\documentclass{emulateapj}
\usepackage{lscape}
\usepackage{epsfig}
\submitted{{\it Accepted for publication in AJ}}

\shortauthors{Bochanski et al.}
\shorttitle{SDSS M Dwarf Spectral Templates}
\begin{document}

\title{Low-Mass Dwarf Template Spectra from the Sloan Digital Sky Survey}

\author{John J. Bochanski\altaffilmark{1},
	Andrew A. West\altaffilmark{1,2},
	 Suzanne L. Hawley\altaffilmark{1},
	 Kevin R. Covey\altaffilmark{1}}

\altaffiltext{1}{Astronomy Department, University of Washington,
   Box 351580, Seattle, WA  98195\\
email: bochansk@astro.washington.edu}
\altaffiltext{2}{Astronomy Department, University of California,
  601 Campbell Hall, Berkeley, CA 94720-3411}

\begin{abstract}
We present template spectra of low-mass (M0-L0) dwarfs derived from over 4,000
Sloan Digital Sky Survey (SDSS) spectra.  These composite spectra are
suitable for use as medium-resolution (R $\sim$ 1,800) radial velocity standards.
We report mean spectral properties (molecular bandhead strengths, equivalent widths) 
and use the templates to investigate
the effects of magnetic activity and metallicity on the spectroscopic and
photometric properties of low-mass stars.  
\end{abstract}

\keywords{\
stars: low mass ---
stars: fundamental parameters ---
stars: M dwarfs ---
stars: activity}

\section{Introduction}

Low-mass dwarfs are the dominant stellar component of the Galaxy.
These ubiquitous stars, with main sequence lifetimes greater than the Hubble time \citep{1997ApJ...482..420L},
have been employed in a variety of Galactic studies: tracing Galactic Disk kinematics
\citep{1996AJ....112.2799H,2002AJ....123.3356G,2003AJ....125.1598L,2005AJ....130.1871B},
describing age-velocity dispersion relations \citep{2006astro.ph..9001W}, and studying Galactic structure
components \citep{1997PASP..109..559R,2001A&A...365..424K,2005ApJ...622..319P}.  Modelling 
of their internal structure,
atmospheric properties and magnetic activity \citep{1993ApJ...406..158B, 1998A&A...337..403B, 1997ARA&A..35..137A, 1999ApJ...512..377H, 2006ApJ...644..484A, 2006astro.ph..9001W} presents interesting theoretical
problems.
Observationally, the spectra of these stars are marked by the presence of 
strong molecular absorption features,
particularly titanium oxide (TiO), which dominates the optical opacity of their cool atmospheres.  The TiO
features, along with molecular bandheads introduced by vanadium oxide (VO) and calcium
hydride (CaH) are used to define the widely accepted M spectral subtype classification scheme
\citep{1991ApJS...77..417K, 1995AJ....110.1838R,1999ApJ...519..802K}.

In order to increase the utility of low-mass stars in large studies of galactic
structure and dynamics, we have been engaged in analyzing spectroscopic data from the
Sloan Digital Sky Survey (SDSS;  \citealp{2000AJ....120.1579Y}), resulting in a series of papers that
describe our methods for photometrically selecting and spectral typing these objects
\citep{2002AJ....123.3409H, 2004AJ....128..426W,2004PASP..116.1105W, 2005PASP..117..706W}
 and discussing their magnetic activity properties
\citep{2004AJ....128..426W,2006astro.ph..9001W}.  
However, we have been unable to exploit the velocity information
contained in the spectra due to the inability of the SDSS pipeline reductions to
provide accurate velocites for M dwarfs \citep{2004AJ....128..502A}.  Our motivation for the present
work is the desire to produce a set of radial velocity templates by combining native,
high-quality, SDSS spectra at each spectral subtype in the M dwarf sequence.  
Additionally, we split the spectra at each subtype into active and inactive
stars, and examine the spectroscopic and photometric properties of these
templates separately, to determine whether the activity is imprinting a signature that
may affect our velocity analysis, and to followup on previous suggestions
that colors and detailed absorption features may change depending on activity level
\citep{1996AJ....112.2799H,1997A&A...319..967A,1999AJ....117.1341H}.

The radial velocity (RV) of an object is the projection of its intrinsic motion onto the line of sight of an observer.
In order to accurately determine the RV of a given star, one must carefully address 
the systematics imposed by the time and location of the observation.  This is usually
accomplished by shifting the frame of the observer to a heliocentric (sun-centered) or barycentric
(center-of-mass centered) rest frame.  The standard method of determining stellar
and galactic RVs has been cross-correlation, as introduced by \cite{1979AJ.....84.1511T}.
This method compares the spectrum of a science target against a known template, using
cross-correlation to determine
the wavelength shift (and therefore velocity) necessary to align the target with the template.
Thus, correlating with a template spectrum that is similar to the science target in all ways
except velocity ensures the most accurate determination of the RV.

 In the following sections, we report on our efforts to establish a set of low-mass star template 
 spectra\footnote{Available at http://www.astro.washington.edu/slh/templates/}
suitable for RV analysis using SDSS spectra at medium (R $\sim$ 1,800) resolution.  In \S \ref{sec:data}, we describe
the observational material from the SDSS
and introduce the problems with the RVs reported for low-mass stars by the standard SDSS spectroscopic pipeline.  
The observations were spectral-typed, inspected for signs of chromospheric
activity, and coadded to form templates at each spectral type, as discussed in \S \ref{sec:analysis}.  The resulting spectral templates, their
accuracy as RV standards, and their spectral and photometric
properties are detailed in \S \ref{sec:results}.  Our conclusions follow in \S \ref{sec:conclusions}.

\section{Data}\label{sec:data}
\subsection{SDSS Photometry}
The Sloan Digital Sky Survey \citep{2000AJ....120.1579Y, 1998AJ....116.3040G,
1996AJ....111.1748F,2001AJ....122.2129H,2002AJ....123.2121S,2002AJ....123..485S,
2003AJ....126.2081A,2003AJ....125.1559P, 2004AJ....128..502A,2004AN....325..583I,2005AJ....129.1755A,2006ApJS..162...38A,2006AJ....131.2332G}
has revolutionized optical astronomy.  Centered on the Northern Galactic Cap,
SDSS has photometrically imaged $\sim$ 8,000 sq. deg. in five filters ($u,g,r,i,z$)
to a faint limit of 22.2 mag in $r$.
This has resulted in photometry of $\sim$ 180
million objects with typical photometric uncertainties of $\sim$ 2\% at $r \sim 20$ \citep{2003MmSAI..74..978I,2006ApJS..162...38A}.
SDSS imaging has been invaluable in recent studies concerning low-mass dwarfs, particularly the
colors of the M star sequence \citep{2004PASP..116.1105W,2005PASP..117..706W} and the study of L and T spectral types
\citep{1999ApJ...522L..61S,2000AJ....119..928F,2000ApJ...536L..35L,2000ApJ...531L..61T,2002AJ....123.3409H,2004AJ....127.3553K,2006AJ....131.2722C}.

\subsection{SDSS Spectroscopy}\label{sec:sdssspec}
Photometry acquired in SDSS imaging mode is used to select spectroscopic followup targets.  The photometry
is analyzed by a host of targeting algorithms (originally described in \citealp{2002AJ....123..485S}) with the primary
spectroscopic targets being galaxies \citep{2002AJ....124.1810S}, luminous red galaxies with z $\sim 0.5 - 1.0$ \citep{2001AJ....122.2267E}, and high redshift quasars \citep{2002AJ....123.2945R}.
Designed to acquire redshifts for $\sim$ 1,000,000 galaxies and 100,000 quasars, twin fiber-fed spectrographs deliver
640 flux-calibrated spectra per 3$^{\circ}$ diameter plate over a wavelength range of 3800-9200\AA ~with a
resolution $R \approx$ 1,800.  Typical observations are the coadded result of multiple 15 minute
exposures, with observations continuing until the signal-to-noise 
ratio per pixel is $>$ 4 at $g = 20.2$ and $i = 19.9$ \citep{2002AJ....123..485S}.  Typically, this takes about 3 exposures.
Wavelength calibrations, good to 5 km s$^{-1}$ or better \citep{2006ApJS..162...38A}, are carried out as described in \cite{2002AJ....123..485S}.
The spectra are then flux-calibrated using F subdwarf standards, with broadband uncertainties
of 4\% \citep{2004AJ....128..502A}.
These observations, obtained and reduced in a uniform manner, form a homogeneous, statistically robust
dataset of over 673,000 spectra \citep{2006ApJS..162...38A}.  SDSS has already
proven to be an excellent source of low-mass stellar spectroscopy
\citep{2002AJ....123.3409H,2003AJ....125.2621R,2004AJ....128..426W,2006AJ....131.1674S}.  Unfortunately, 
the RVs determined for low-mass stars in the SDSS pipeline are known to be inaccurate\footnote{See http://www.sdss.org/dr5/products/spectra/radvelocity.html}\citep{2004AJ....128..502A}.  These systematic errors, on the 
order of 10 km s$^{-1}$ \citep{2004AJ....128..502A} result
primarily from spectral mismatch, since there are only four low-mass stellar templates in the standard spectroscopic pipeline.  Thus, we sought to remedy this situation by establishing
a uniform set of low-mass RV templates derived from SDSS spectroscopy.

\section{Analysis}\label{sec:analysis}
To build a database of low-mass stellar spectra, we queried the Data Release 3
(DR3; \citealp{2005AJ....129.1755A}) Catalog Archive Server (CAS)
for spectra with late-type dwarf colors (from \citealp{2005PASP..117..706W}), using $0.5 < r-i < 3.05$ and $0.3 < i-z < 1.9$.
The color ranges quoted in \cite{2005PASP..117..706W} were slightly extended to increase the total
number of low-mass stellar spectra.
These color cuts were the only restrictions applied to the DR3 data.  We treated each spectral 
subtype independently, performing 11 (M0-L0) queries, some of which overlapped in color-color space (see Table  \ref{table:colorcuts}).
Thus, some spectra were selected twice, usually in neighboring spectral subtypes (i.e., M0 and M1).  
These queries yielded $\sim$~133,000 candidate spectra in the 11 (M0-L0) spectral type bins.

\subsection{Spectral Types and Activity}
The candidate spectra were examined with a suite of software specifically designed to
analyze M dwarf spectra.  This pipeline, as introduced in \cite{2002AJ....123.3409H},
measures a host of molecular band indices (TiO2, TiO3, TiO4, TiO5, TiO8, CaH1, CaH2, and CaH3),
and employs relations first described by \cite{1995AJ....110.1838R} to determine a spectral type
from the strength of the TiO5 bandhead.
All spectral types were confirmed by manual inspection, adjusting the final spectral type, if 
necessary.  The accuracy of the final spectral type
is $\pm 1$ subtype.  Additionally, the software pipeline measures
the equivalent width (EW) of the H$\alpha$ line, quantifying the level of magnetic activity in a given
low-mass dwarf. See \cite{2004AJ....128..426W} for details on the measurement of bandheads 
and line strengths in low-mass star SDSS spectra.

Inspection of each candidate spectrum allowed us to remove contaminants (mostly galaxies)
from the sample, reducing its size to $\sim$20,000 stellar spectra.  We also removed the white dwarf-M dwarf
pairs that were photometrically identified by \cite{2004ApJ...615L.141S}.
The database was then culled of duplicates.  As shown in \cite{2005PASP..117..706W} (and Table \ref{table:colorcuts}), 
M dwarfs
of different spectral types can possess similar SDSS photometric colors.  Thus, some spectra
with 
overlapping photometric colors were duplicated in our original database (see \S \ref{sec:analysis} and Table \ref{table:colorcuts}).  
Each duplicate spectrum was identified by filename and in cases where different 
spectral types were manually assigned by eye to the same star (typically one subtype apart), the earlier spectral type was kept. 
The typical difference in spectral type was 
one subclass, in agreement with our stated accuracy.  These various cuts reduced
the sample from $\sim$20,000 to $\sim$12,000 stars.   

The spectra were then categorized based on their activity.
In order to be considered active, a star had to meet the criteria originally
described in \cite{2004AJ....128..426W}: (1) The measured H$\alpha$ EW is larger than
1.0 \AA;  (2) The measured EW is larger than the error; (3)  The height of the H$\alpha$
line must be three times the noise at line center; (4) The measured EW
must be larger than the average EW in two 50 \AA\ comparison regions
(6500-6550 \AA\ and 6575-6625 \AA).  In order to be considered inactive, the measured
EW had to be less than 1.0 \AA\  and have a signal-to-noise ratio greater than three in the comparison regions.
By selecting only these active and inactive stars, our final sample is limited to spectra
with well-measured features, removing spectra with low signal-to-noise ratios.  The 
resulting database consisted of $\sim$6,000 stellar spectra.

\subsection{Coaddition}
SDSS spectra are corrected to a heliocentric rest frame during the standard pipeline 
reduction and are on a vacuum wavelength scale.  
To assemble the fiducial template spectra,
stars of a given spectral type were first shifted to a zero-velocity rest frame, then normalized and coadded.
Multiple strong spectral
lines were used to measure the velocity of each star to obtain an accurate shift to the rest frame.
For inactive stars, the red line (7699 \AA) of the K {\rm I} doublet and both lines of the
Na {\rm I} doublet (8183, 8195 \AA) were measured, while in active stars, H$\alpha$
was also used.  These spectral line combinations were selected for their strength in all
low-mass stellar spectra, ensuring that no single line would determine the final velocity of a star.
Spectral lines were fit with single Gaussians and inspected visually
to ensure proper fitting.  Stars with spurious fits or discrepant line velocities 
(lines which deviated from the mean by $>$ 30 km s$^{-1}$) 
were removed from the
final coaddition.  Removing these spectra reduced the final sample size to $\sim$ 4,300 stars.
The classical redshift correction was then applied to each spectrum in the final sample, justifying them
to a zero-velocity rest frame.  

The velocity of each spectral line in the final sample was fit with an 
measurement uncertainty of $\sim$ 10 km s$^{-1}$, as determined
by the mean scatter among individual line RV measurements.  
In wavelength space, this translates to about 0.2 \AA\ resolution 
near H$\alpha$ (note the resolution of SDSS (R $\sim$ 1,800) implies 3.6 \AA\ resolution at H$\alpha$).  
Since the observed spectrum is a discretization of a continuous 
flux source (i.e., the star), wavelength shifts introduced by the radial velocity of a star will move 
flux within and between resolution elements.  These wavelength shifts, which are resolved to
subpixel accuracy, act to increase the resolution of the final co-added spectrum (see 
\citealp{1996OptEn..35.1503P}).  This is similar to the common ``drizzle'' technique of 
using multiple, spatially distinct low-resolution
images to produce a single high-resolution image \citep{2002PASP..114..144F}.

\begin{figure*}[htbp]
    \centering
    \includegraphics[scale=0.85]{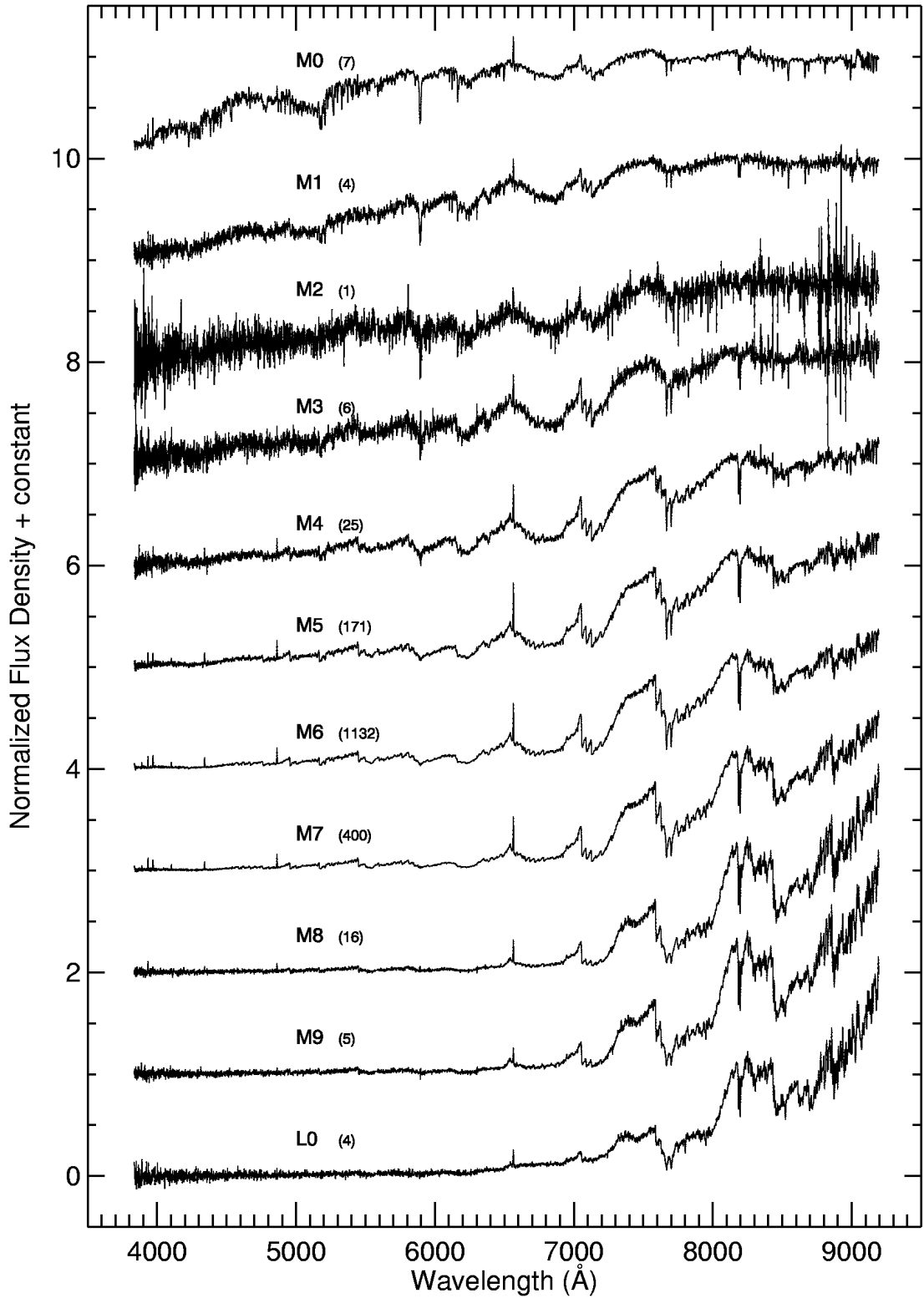}
    \caption{Mean template spectra of active low-mass dwarfs of types M0-L0.  The spectral type
    and number of stars (in parentheses) are labeled for each template.}
\label{fig:active}
\end{figure*}

\begin{figure*}[htbp]
    \centering
    \includegraphics[scale=0.85]{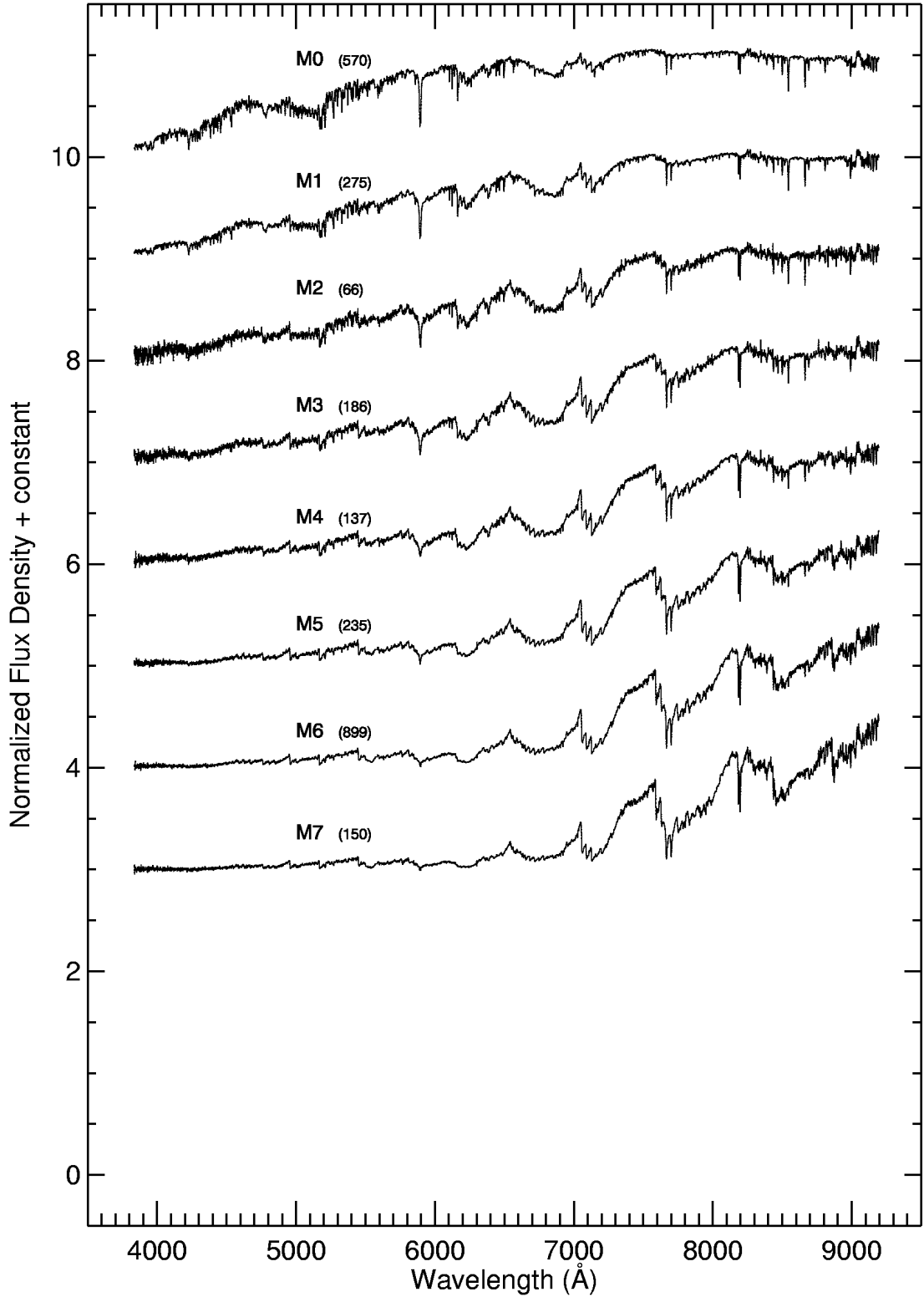}
    \caption{Mean template spectra of inactive low-mass dwarfs of types M0-M7. No M8-L0 dwarfs met the
    activity and consistent line velocity criteria. Spectral type
    and number of stars in each template are labeled.}
\label{fig:inactive}
\end{figure*}

\begin{figure*}[htbp]
    \centering
    \includegraphics[scale=0.85]{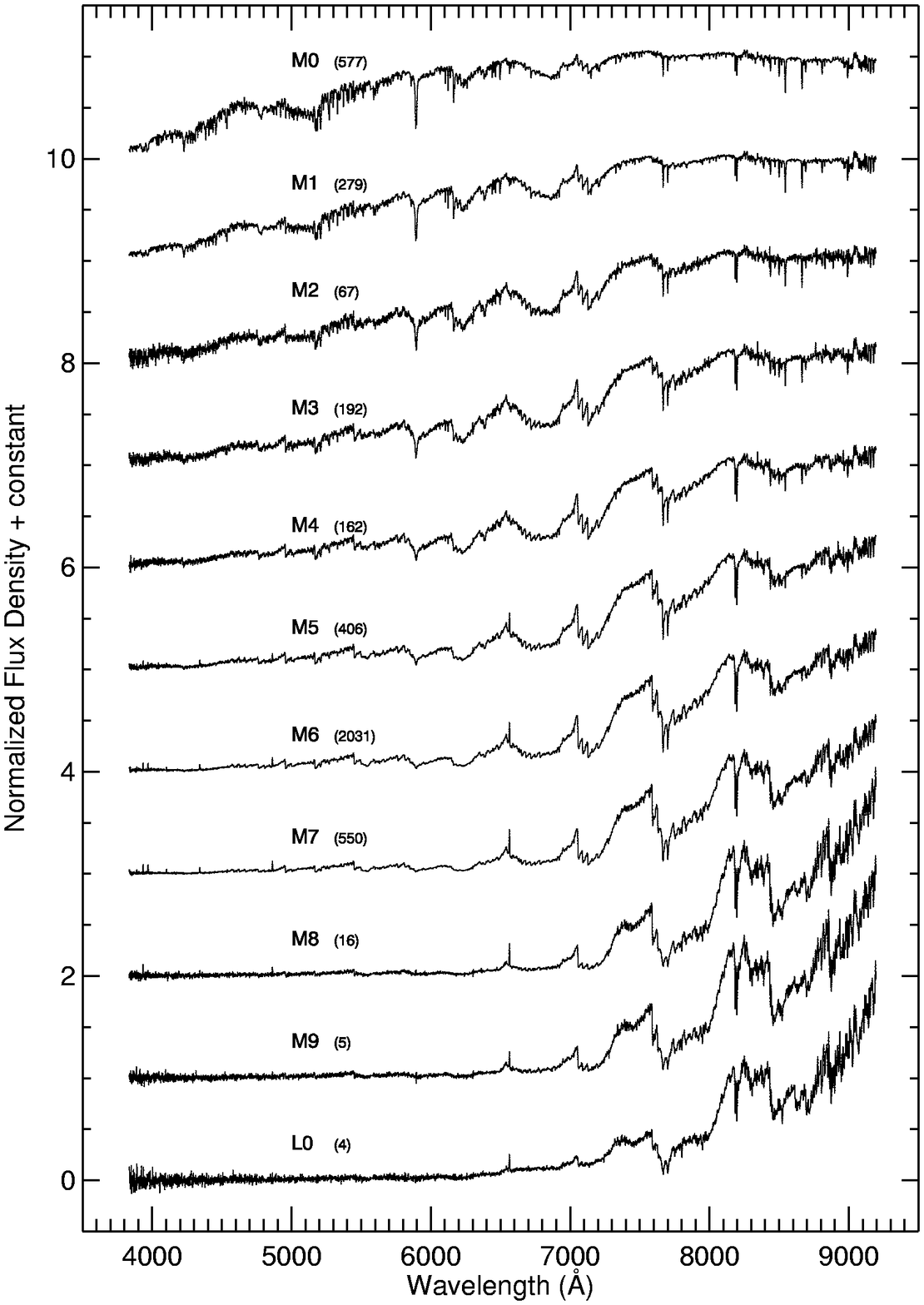}
    \caption{Mean composite template spectra for low-mass dwarfs of types M0-L0, formed by
    combining the data shown in Figures 1 and 2. Spectral type
    and number of stars in each template are labeled.}
\label{fig:all}
\end{figure*}

The wavelength-justified spectra were then normalized at 8350\AA\ and coadded (with equal
weighting of each spectrum) using the following prescription.
At each subtype, we attempted to construct three templates: one composed solely of
active stars (Fig. \ref{fig:active}), one composed of inactive stars (Fig. \ref{fig:inactive}),
and a third composed of both the inactive and active stars
from the previous two sets (Fig \ref{fig:all}).  For each template,
the mean and standard deviation were calculated at each pixel.  In later ($>$ M7) subtypes, the lack of
spectra meeting our activity and velocity criteria resulted in fewer templates.

\section{Results and Discussion}\label{sec:results}
\begin{figure}[htbp]
    \centering
    \includegraphics[scale=0.5]{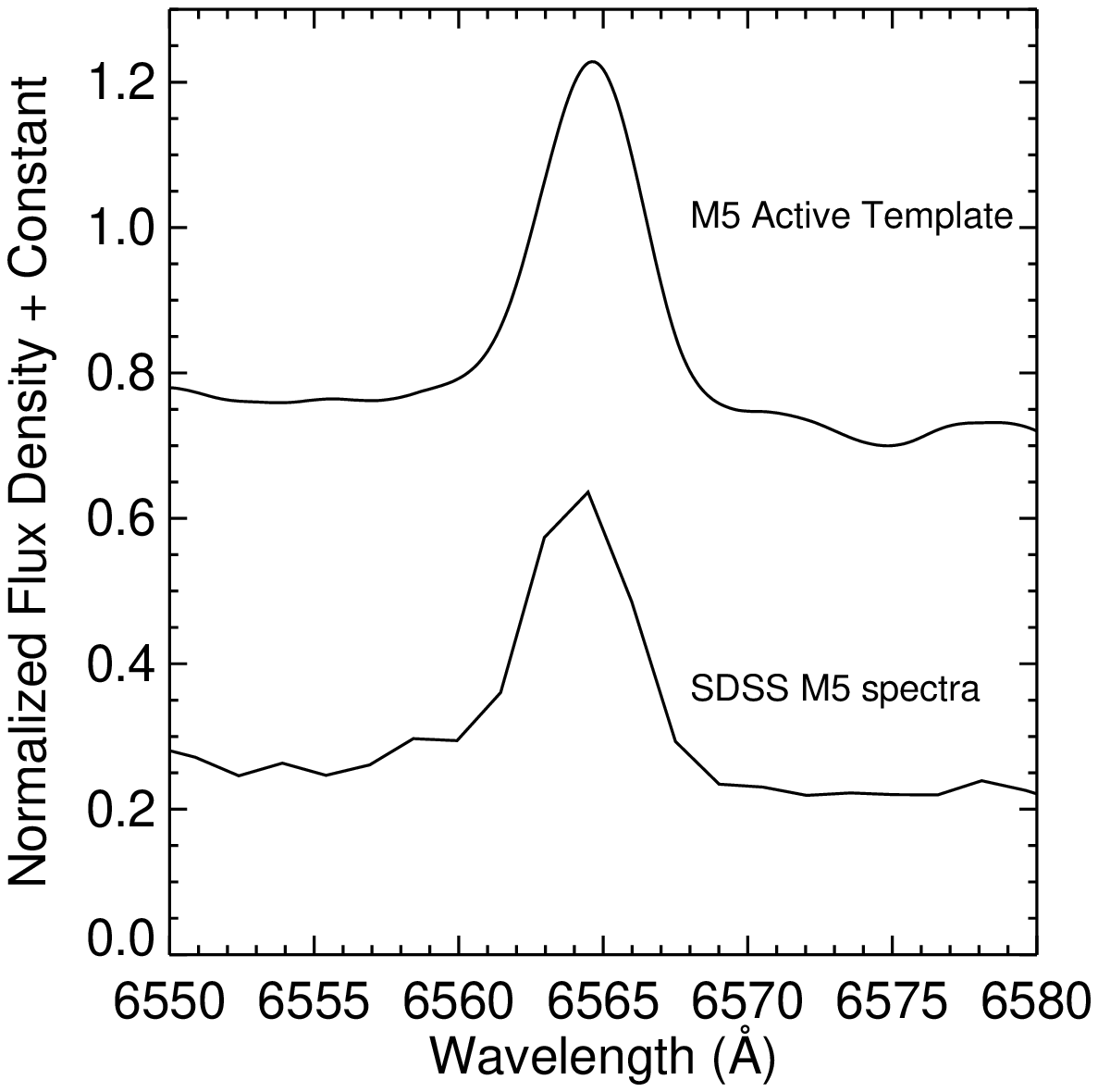}
    \caption{Comparison of a high signal-to-noise SDSS M5 spectrum (bottom) to the active M5 template (top) near H$\alpha$.  Note
    the higher resolution of the template spectrum.}
\label{fig:hires}
\end{figure}

The final template spectra (Figures \ref{fig:active}-\ref{fig:all}) represent
the mean spectral properties of low-mass dwarfs as observed by the SDSS spectrographs.
In Figure \ref{fig:spec}, we show illustrative examples of our templates for an 
inactive M1 star and an active M6 star, with
strong atomic lines and molecular bandheads labeled.  Prominent molecules include MgH, CaH, TiO, VO and CaOH.  The active stars show the Balmer series to H8 ($\lambda \sim$ 3889 \AA)
along with Ca {\rm II} H and K ($\lambda \sim$ 3968, 3933 \AA).  In Figure \ref{fig:hires} we compare 
a high signal-to-noise SDSS spectrum of an M5 star to its template counterpart in the region near H$\alpha$.  
It is clear that the template
has significantly higher spectral resolution; e.g. a weak feature near 6575 \AA\ is visible only in the template.  
In the following sections, we 
explore the feasibility of using these templates
 as RV standards and the effects of chromospheric activity and metallicity on the mean spectroscopic and photometric
 characteristics of low-mass stars.
 
 \subsection{Radial Velocity Accuracy}
The primary uncertainties associated with determining
RVs using the cross-correlation method are due to the resolution of the spectra, accuracy of the wavelength 
calibration and matching the spectral type of the template and science data. To ensure that the RVs 
measured with our templates are accurate, we have carried out
tests that quantify the internal consistency and external zero-point precision of these templates.  These
tests are described below.

\subsubsection{Internal Consistency}
To quantify the internal consistency among templates, sequential spectral types were
cross-correlated using the $fxcor$ task in IRAF\footnote{
IRAF is distributed by the National Optical Astronomy Observatories,
which are operated by the Association of Universities for Research
in Astronomy, Inc., under cooperative agreement with the National
Science Foundation.}.  This minimizes the error introduced by spectral type mismatch, which often dominates
the errors associated with cross-correlation redshift measurements \citep{1979AJ.....84.1511T}.  Thus,
wavelength calibration and intrinsic resolution are the major sources 
of uncertainty in our analysis.  In all cases (i.e., active, inactive and
combined templates) the mean difference in velocity between adjacent spectral subtypes 
was $\lesssim$ 1 km s$^{-1}$.
This compares favorably with the 3.5 km s$^{-1}$ spread in SDSS data as reported by \cite{2000AJ....120.1579Y}.
Note this value is derived from observations of stars in M67 \citep{1986AJ.....92.1100M}, and does not include any low-mass dwarfs.
 

 \begin{figure*}[htbp]
    \centering
    \plottwo{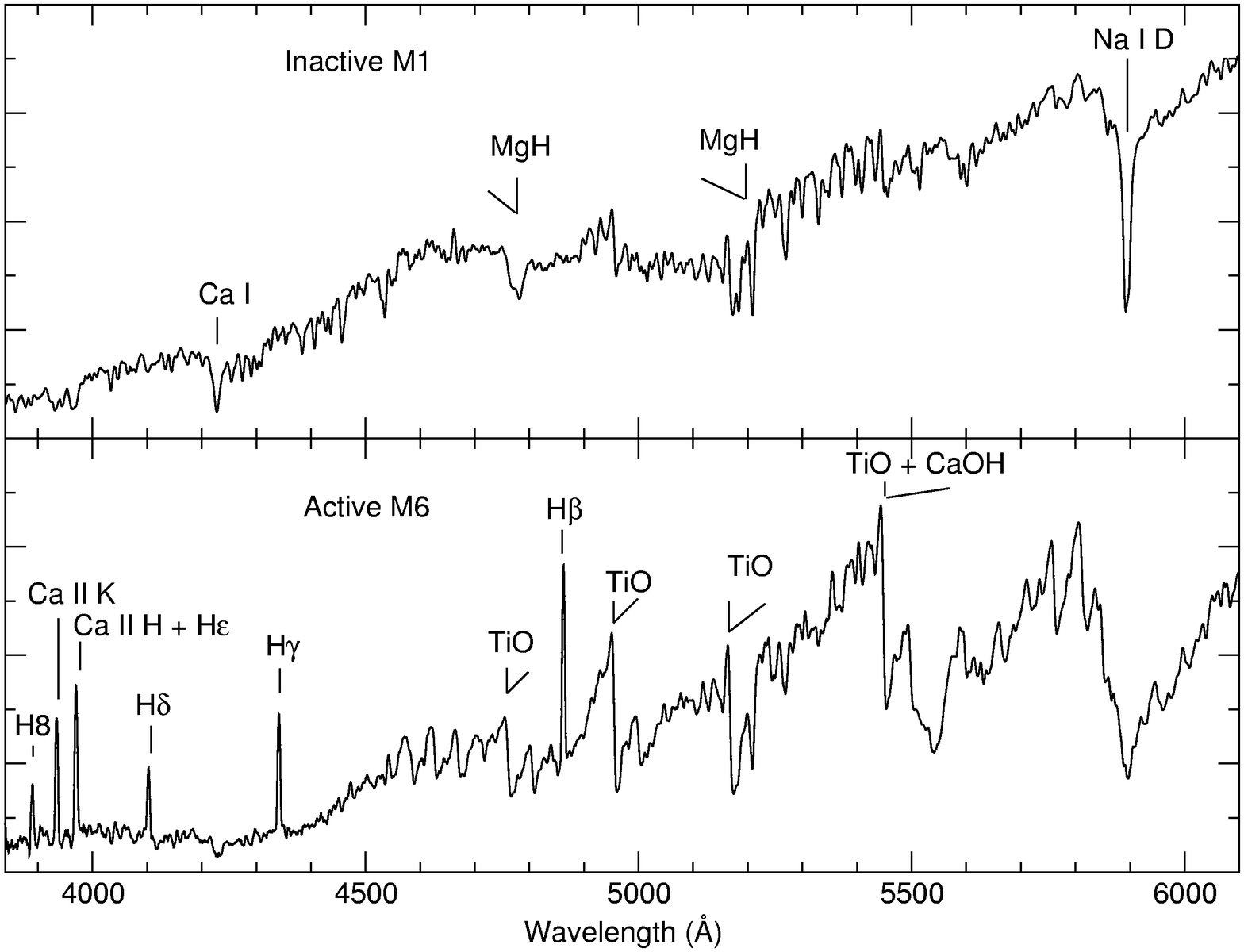}{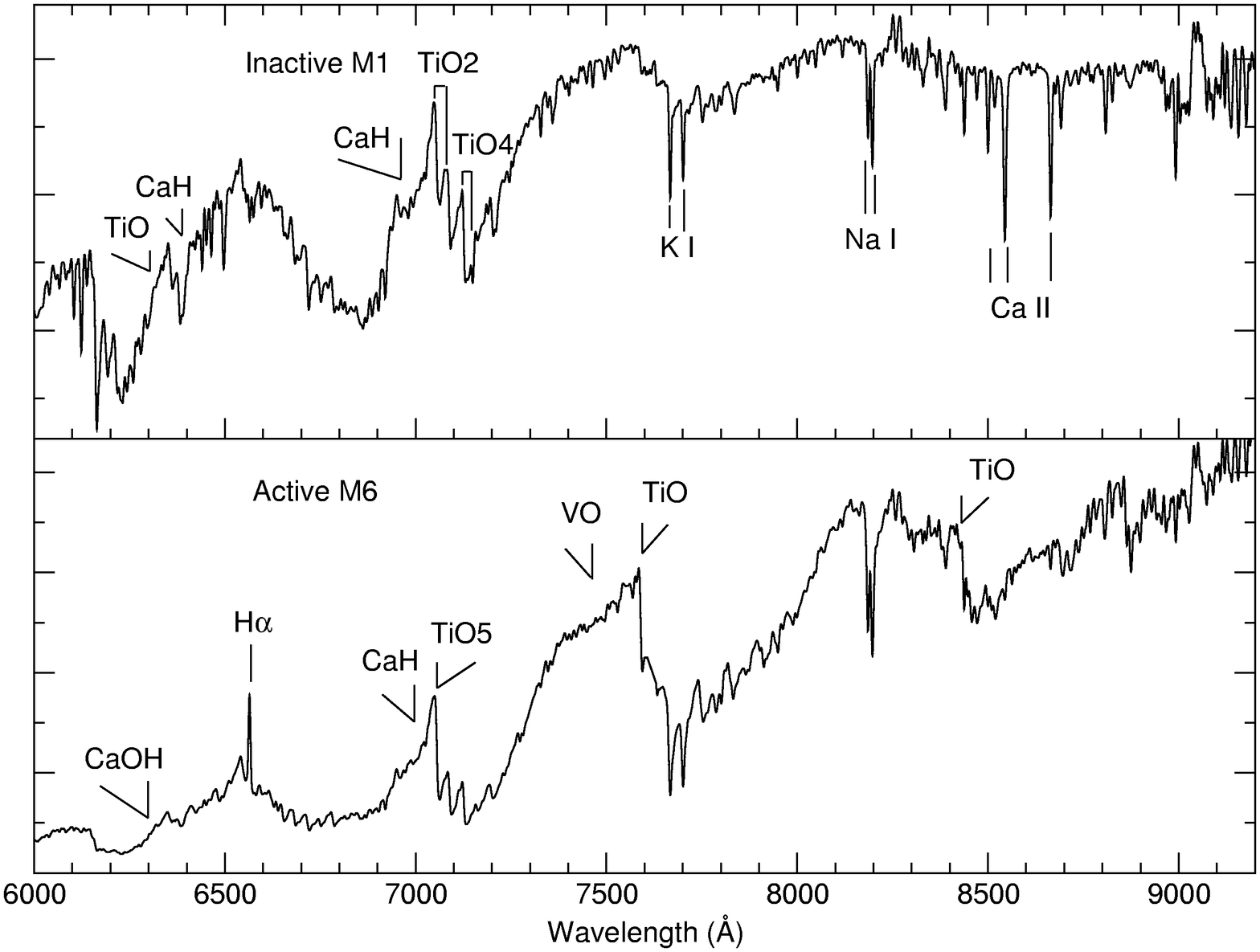}
    \caption{Illustrative template spectra of an inactive M1 star and an active M6 star with strong
molecular and atomic features labeled.}
\label{fig:spec}
\end{figure*}

\subsubsection{External Consistency: Hyades}
To test the external accuracy of the template spectra, they were cross-correlated
against Hyades cluster members with well-measured RVs, observed as
part of our SDSS collaboration effort to produce RV standards for low-mass dwarfs.
Each Hyades star has a known RV
\citep{2000MNRAS.316..827R,1994AJ....108..160S,1997ApJ...475..604S,2000AJ....119.1303T,1988AJ.....96..172G}
or is a confirmed member of the cluster, whose dispersion is $<$ 1 km s$^{-1}$ \citep{1988AJ.....96..198G,2000A&A...358..923M}.  
The Hyades RVs in the literature were measured from high-resolution echelle spectra, with a typical accuracy of
$\lesssim$ 1 km s$^{-1}$.  Thus, the SDSS spectra of these Hyads provide a way to check our cross-correlation RVs 
against an external standard system. 

\begin{figure}[htbp]
 \centering
    \includegraphics[scale=0.45]{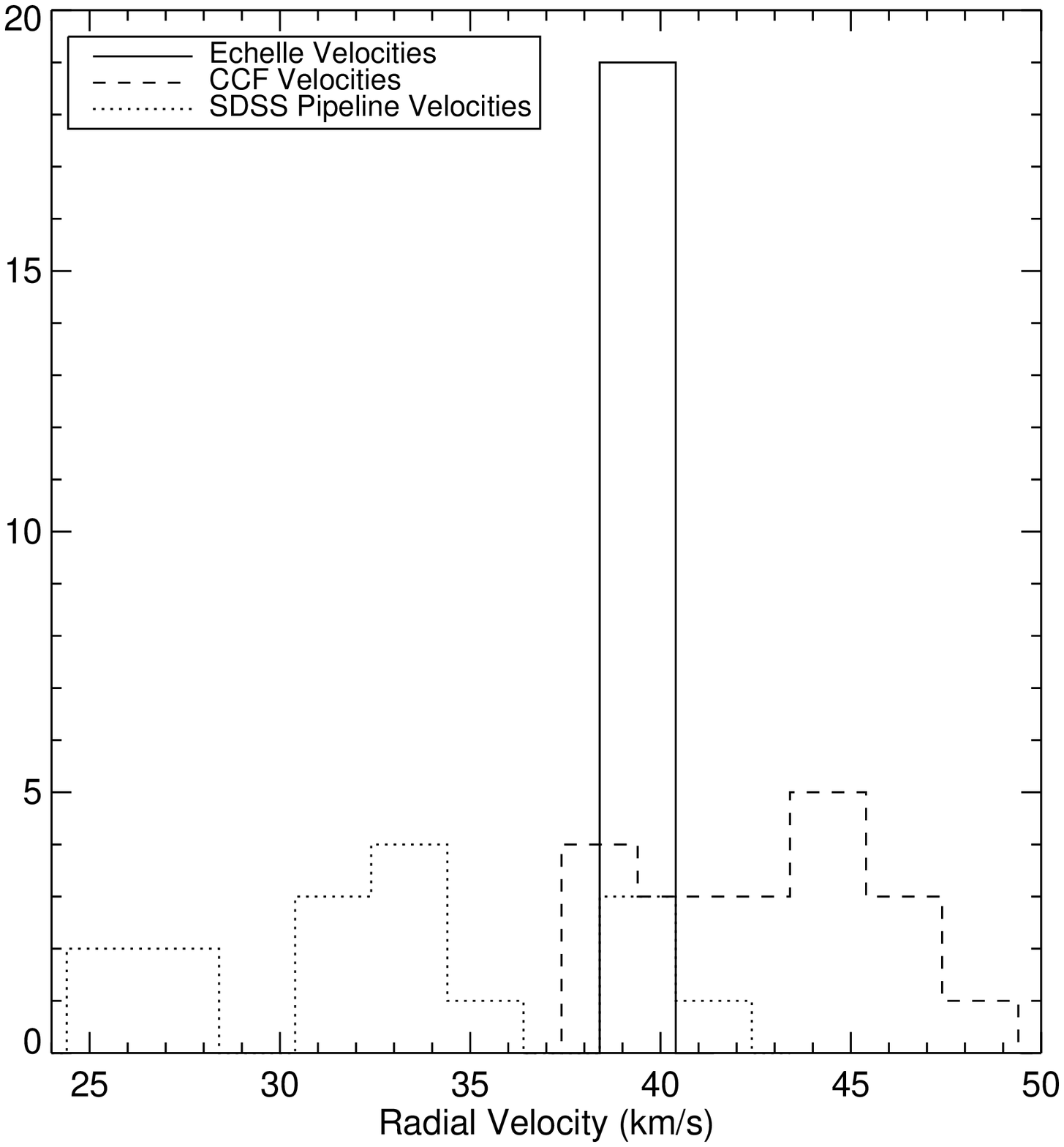}
    \caption{Histogram comparison of high-resolution echelle RV measurements (solid line), medium resolution
    cross-correlation RV determinations (dashed line), and SDSS pipeline RVs (dotted line) for 19 Hyades M dwarfs.  Note the SDSS pipeline  
    reported wildly discrepant velocities for two stars (not shown).}
\label{fig:rv_comp}
\end{figure}

The medium-resolution spectra of the Hyades stars secured by SDSS were correlated 
against our templates, with results shown in Figure \ref{fig:rv_comp}.  The 
high-resolution echelle data have a mean of 38.8 km s$^{-1}$ and a standard deviation of 0.27 km s$^{-1}$.
The RVs measured with our template spectra yield a mean RV of 42.6 km s$^{-1}$ with a standard
deviation of 3.2 km s$^{-1}$.  By comparison, the SDSS pipeline RVs produced a mean velocity of 31.9 km s$^{-1}$ and a 
standard deviation of 6.8 km s$^{-1}$ (after removing two highly discrepant measurements).
Using the template spectra better reproduces the coherent velocity signature of 
the Hyades and provides much more reliable velocities than the standard SDSS pipeline measurements.  The templates
are therefore well-suited for use as medium-resolution RV standards for low-mass dwarfs.


\subsection{Spectral Differences: Activity \& Metallicity}
The effect of magnetic activity on the spectral properties of a low-mass star is clearly
manifested by the existence of emission lines.  This effect is often quantified by measuring
the luminosity in the H$\alpha$ line divided by the bolometric luminosity ($L_{\rm H\alpha}/L_{\rm Bol}$).
Other changes due to activity, such as varying strength of molecular
bandheads \citep{1996AJ....112.2799H, 1999AJ....117.1341H} and changes in the shape of the
continuum have been sparsely investigated.  Additionally, metallicity affects the strength of 
molecular bandheads at a given temperature (see \citealp{2006PASP..118..218W}).
Using the template spectra as fiducial examples of thin-disk, solar-metallicity low-mass stars, 
we next examine changes in the spectral properties of low-mass stars introduced by magnetic 
activity and metallicity.

\subsubsection{Activity: Decrements \& $L_{\rm H\alpha}/L_{\rm Bol}$}
Emission features are dependent on the temperature and density structure of the outer stellar atmosphere.
The line fluxes of the Balmer series lines and the Ca {\rm II} K line ($\lambda \sim 3933$ \AA) can be
used to examine the structure of the chromosphere in magnetically active stars
\citep{1995MNRAS.272..828R,2006PASP..118..617R} and to investigate chromospheric
heating in quiescent (i.e. non-flaring) dMe stars \citep{1994A&A...281..129M, 1997A&A...326..249M}.  The
Balmer decrement (ratios of Balmer line strengths to a fiducial, here
taken to be H$\beta$) is traditionally used to quantify medium-resolution
spectra.  Table \ref{table:decrements} gives the Balmer series and 
Ca {\rm II} K decrements for the active templates.  
There is no strong trend in the decrement with spectral type for the 
template spectra, which 
is consistent with the previous study of \cite{1989A&A...217..187P}, who
reported average decrements
over a range of K and M spectral types.  However, both the templates
and \cite{1989A&A...217..187P} show a gradual increase in the H$\alpha/$H$\beta$ ratio
with spectral type (see Table \ref{table:decrements}).  Evidently the
structure and heating of low-mass stellar chromospheres remains 
fairly similar over the range of M dwarf effective temperature (mass),
with the H$\alpha$ gradually becoming stronger relative to the
higher order Balmer lines at later spectral type.

\begin{figure}[htbp]
  \centering
    \includegraphics[scale=0.6]{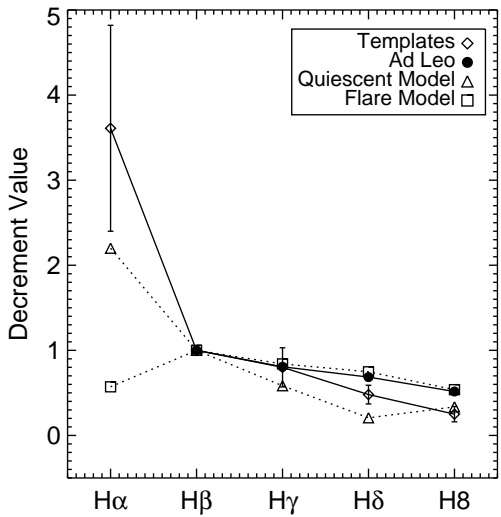}
    \caption{Decrements for the averaged active templates (open diamonds, error bars)
are shown. Also shown are results for the dM3e star AD Leo:
observed flare (filled circles; from \citealp{1991ApJ...378..725H}), model 
flare (open squares), and model quiescent (open triangles) decrements \citep{2006ApJ...644..484A}. 
Flaring atmospheres have increased density at chromospheric temperatures, resulting in higher opacity and
increased emission in the higher-order Balmer lines compared to H$\alpha$.  The result is a relatively
flatter decrement.}
\label{fig:decrements}
\end{figure}

The Balmer decrements observed in 
AD Leo (dM3e) during a large flare 
\citep{1991ApJ...378..725H} and determined for quiescent and 
flaring model atmospheres \citep{2006ApJ...644..484A} are given 
in Table \ref{table:decrements} for comparison. 
In Figure \ref{fig:decrements}, we plot these observed and
model decrements together with the average 
Balmer decrements of the active low-mass stellar templates.  
The flare decrements, both observed and model, are much flatter than
those in the non-flaring atmospheres, suggesting (though within the errors)
increased emission in the higher
order Balmer lines.  
This probably reflects the higher chromospheric densities
(hence greater optical depth in the Balmer line-forming region) 
in the flare atmospheres.  Evidently the range in chromospheric
properties among M dwarfs of different spectral types is much less
(during quiescent periods) than in a given star between its quiet
and flaring behavior.

For completeness, the average Balmer line and Ca II K EWs and the
quantity $L_{\rm H\alpha}/L_{\rm Bol}$ measured from the active templates are reported
in Table \ref{table:ew}.  
$L_{\rm H\alpha}/L_{\rm Bol}$ which is used to quantify activity, was calculated using the H$\alpha$ EW and the ($i-z$) continuum ($\chi$)
relation of \cite{2005PASP..117..706W}, as first described in \cite{2004PASP..116.1105W}.
The average EWs reported in Table \ref{table:ew} agree with previous results \citep{1997ApJ...475..604S, 2002AJ....123.3356G}.
The general increase toward later types is attributed to the lower 
continuum flux in the vicinity of H$\alpha$ as the stellar effective
temperature decreases.  
The $L_{\rm H\alpha}/L_{\rm Bol}$ ratios are also consistent with 
previous studies \citep{1996AJ....112.2799H,2000AJ....120.1085G,2004AJ....128..426W}, 
attaining a relatively constant value (with large scatter)
among early-mid M (M0-M5) types, and decreasing at later types.


\subsubsection{Activity: Bandheads}

Molecular rotational and vibrational transitions imprint large bandheads on the observed spectra
of low-mass stars.  The strength of the TiO bandheads in the visible 
is often used as a spectral-type discriminant
\citep{1995AJ....110.1838R}.  Additionally, CaOH, TiO and CaH bandheads have been employed as temperature and metallicity indicators \citep{1997AJ....113..806G,1999AJ....117.1341H, 2006PASP..118..218W}.  Following previous conventions 
\citep{1995AJ....110.1838R,1999ApJ...519..802K},  we provide measurements
for the CaH bandheads at~$\sim$~6400\AA\ (CaH1), $\sim$~6800\AA\ (CaH2,CaH3) in Table \ref{table:cah}, and for the TiO 
bandheads at $\sim$~7050\AA\ (TiO2,TiO4,TiO5) and $\sim$~8430\AA\ (TiO8) in Table \ref{table:tio}.  
Note that TiO2 and TiO4 are sub-bands of the full TiO5 bandhead.

\begin{figure}[htbp]
    \centering
    \includegraphics[scale=0.55]{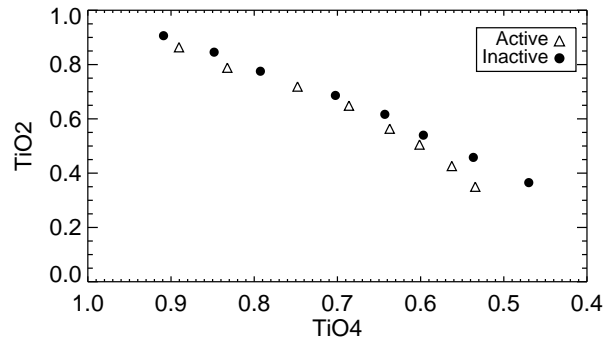}
    \caption{TiO2 vs. TiO4 for active (open triangles) and inactive (filled circles) templates.  At
    a constant TiO4 value, TiO2 is deeper (smaller index values) in the active stars, confirming the result
    originally discussed in \cite{1996AJ....112.2799H}.}
\label{fig:tio2tio4}
\end{figure}

As first observed by \cite{1996AJ....112.2799H}, activity can introduce changes in the TiO bandheads \citep{1999AJ....117.1341H, 1999MNRAS.302...59M}.  
Shown in Figure \ref{fig:tio2tio4} is the TiO2 index as a function of the TiO4 index.  For active stars, 
the strength of the TiO2 bandhead is increased (smaller 
index) at a given value of the TiO4 index.  Alternately, at a given index of TiO2, the TiO4 index is
weaker in active stars.  This provides interesting constraints on the 
structure of the atmosphere, suggesting that the formation of TiO, thought to take place
near the temperature minimum region below the chromospheric
temperature rise 
\citep{2005astro.ph..9798C, 2005nlds.book.....R}, is affected by the presence of an
overlying chromosphere. 
The opposite behavior of these two sub-bands serves to decrease the sensitivity of TiO5 to chromospheric activity, 
making it a good temperature (spectral-type) proxy regardless of the activity level of the star (see \citealp{1999AJ....117.1341H}).


\subsubsection{Activity: Spectral Features}\label{sec:specfeatures}
\begin{figure}[htbp]
    \centering
    \includegraphics[scale=0.3]{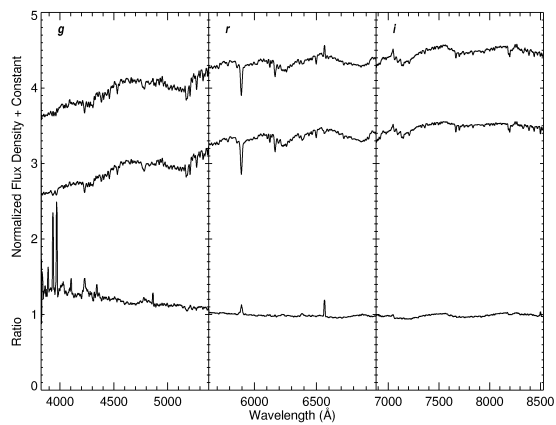}
    \caption{Shown is an illustrative example of our flux-ratio analysis.  The active M0 spectral template (top) is 
 divided by the inactive (middle) template.  The resulting flux ratio is plotted on the bottom.  The windows display the approximate wavelength 
bounds of the SDSS $g$,$r$, and $i$ filters.}
\label{fig:ratios}
\end{figure}

Two main effects influence the colors of active stars:  
the presence of emission lines and changes in the continuum emission.
To investigate these effects, we divided the active 
templates by their inactive counterparts.  
Figure \ref{fig:ratios} is an illustrative example of our analysis,
showing the individual active and inactive M0 template spectra 
together with the ratio of the active to inactive flux.  The
approximate wavelength bounds of the SDSS $g, r$ and $i$ filters 
are indicated. The ratio
shows enhanced blue continuum emission in the active template, and
significantly enhanced emission lines, particularly in Ca II H and K.
These effects lead
to a bluer ($g-r$) color for the active star (see Table \ref{table:colors},
discussed further in \S \ref{sec:photcolors} below).  The change in color is
dominated by the continuum enhancement, with the increased emission
line flux providing only a marginal effect.  Similar continuum
and line flux enhancements are observed during flares, suggesting
that the active M0 template may include one or more stellar spectra
obtained during low level flaring conditions.  As described in \cite{2002ApJ...580L..73G}, 
low level flaring maybe responsible for a significant fraction of the ``quiescent''
chromospheric emission observed on active stars. 

The ratio also shows two ``emission'' lines in the $r$ band corresponding 
to emission in the core of the Na {\rm I} D doublet ($\lambda \sim$5900 \AA, 
doublet marginally resolved at SDSS resolution, but well resolved in our templates) and in H$\alpha$.
Again, these lines do not significantly contribute to the combined 
flux of the template in the $r$ band, 
as shown by the marginally redder $r-i$ color for the active
template in Table \ref{table:colors}.  The spectral flux ratio in
the $i$ band is very close to unity, with no strongly 
varying emission or continuum features between the active and inactive 
templates.  This analysis suggests that changes in the continuum emission of 
active stars provide the most important contribution to observed color 
differences.
 
\begin{figure}[htbp]
    \centering
    \includegraphics[scale=0.4]{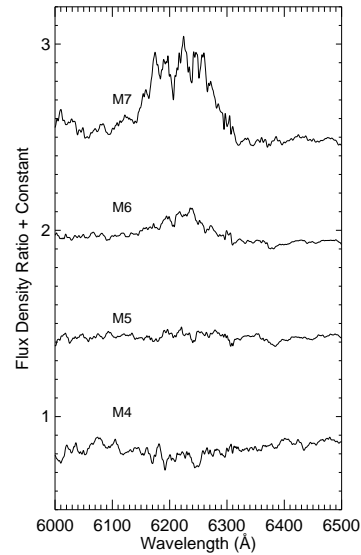}
    \caption{The ratio of the active to inactive flux for the 
M4-M7 templates is shown in the vicinity of the CaOH (6230 \AA) 
bandhead.  Note the increase in the ratio at later types, indicating 
a shallower CaOH band in the active template compared to its inactive 
counterpart.  Apparently, the formation mechanism of CaOH depends on 
both the star's spectral type (effective temperature) and the 
presence of a chromosphere.}

\label{fig:caoh}
\end{figure}

The variable strength of the 
CaOH (6230 \AA) bandhead is also of note.  Shown in 
Figure \ref{fig:caoh} are the active to inactive ratios for M4-M7 subtypes.
The growth of this feature indicates there is some dependence 
of the formation mechanism of CaOH on spectral type (effective
temperature, mass), perhaps
changing the position of the temperature minimum within the atmosphere.
Note that the feature near this wavelength previously discussed
as a good temperature indicator by \cite{1999AJ....117.1341H}
is actually due to TiO in early M dwarfs.  CaOH begins to dominate
the opacity in this region only at types later than M4, which were
not available to observation in the clusters described by
\cite{1999AJ....117.1341H}.  Therefore these new SDSS observations
are the first evidence of a real effect in the CaOH band that
differs with the presence of a chromosphere and changes with
spectral type.  These observations,
together with the differences in the TiO2 and TiO4 bands should
provide strong constraints on the next generation of atmospheric
models (including chromospheres) for M dwarfs.


\subsubsection{Metallicity: Spectral Features}
\begin{figure*}[htbp]
    \centering

    \plottwo{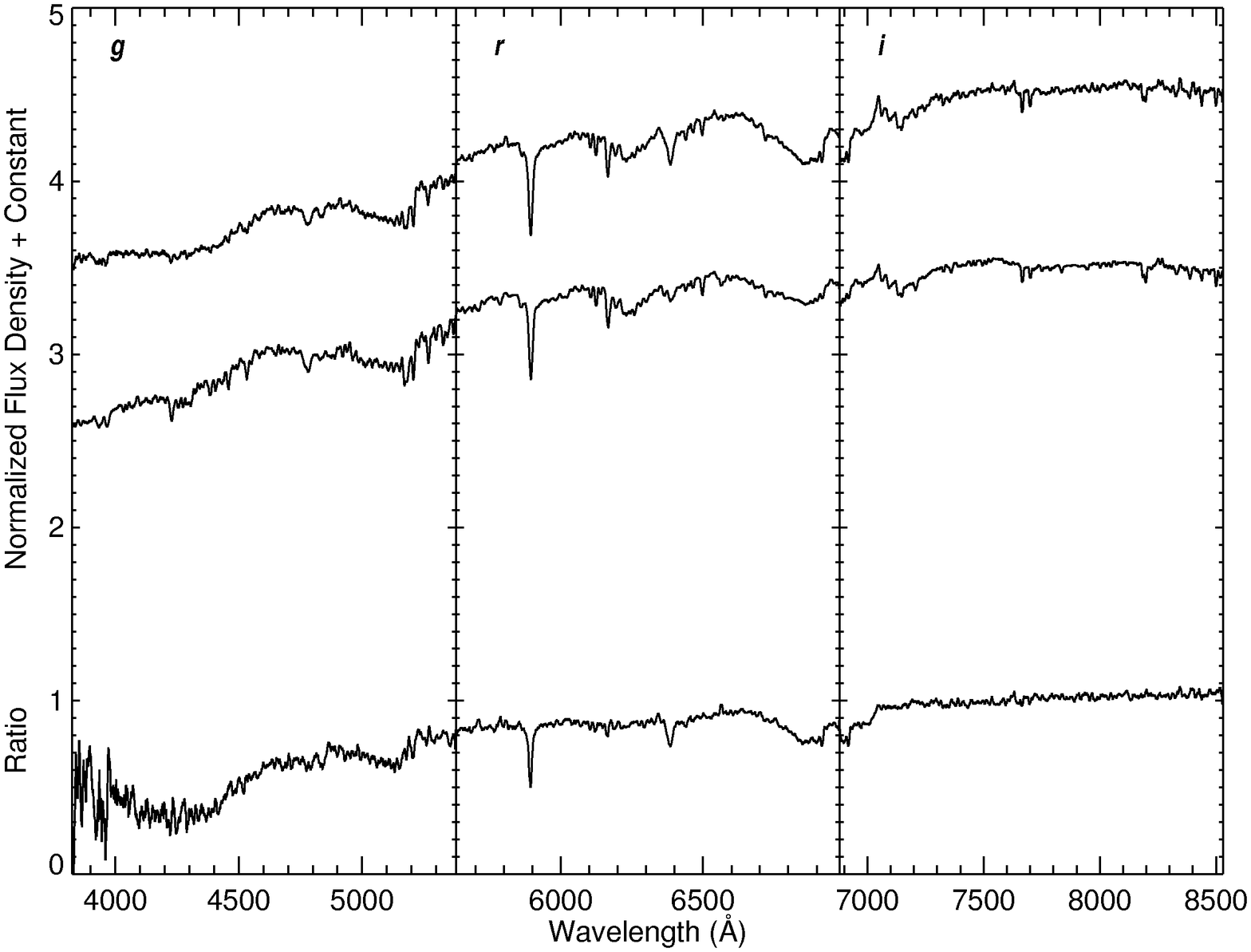}{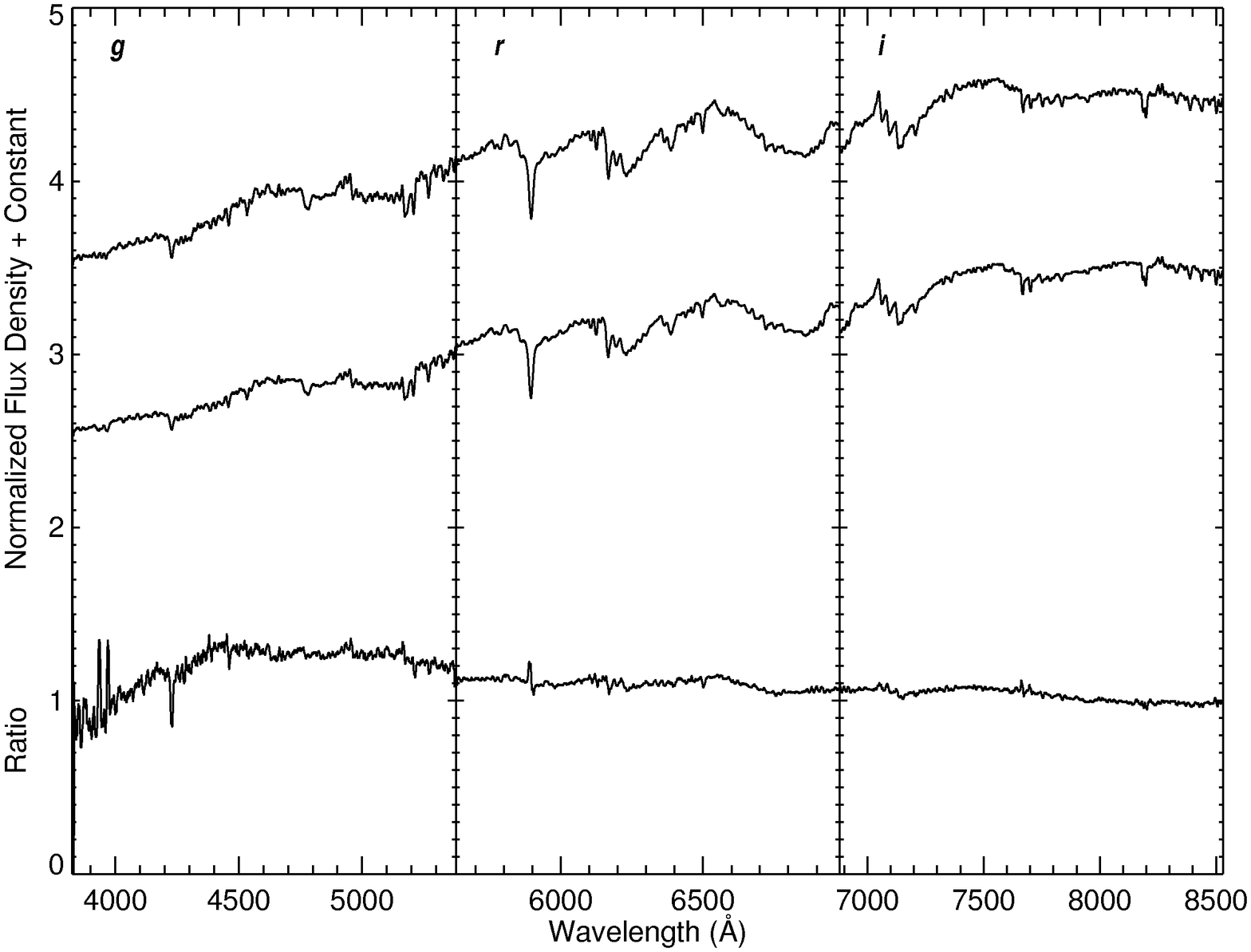}
    \caption{Shown are illustrative examples of our flux ratio analysis 
of an M0 subdwarf (left panel) and an M1 Hyades star (right panel).  
The subdwarf and Hyades star are the top spectra in their respective panels; the 
composite template of the same spectral type (from Figure \ref{fig:all}) appears
as the middle spectrum, and the ratio of the two spectra is shown on 
the bottom.  }
\label{fig:metalratios}
\end{figure*}

We explored the effects of metallicity on the spectra by comparing 
our composite templates to a low-metallicity subdwarf ([Fe/H] $\sim -0.5$; \citealp{2006PASP..118..218W})  and a 
metal-rich Hyades dwarf ([Fe/H] = 0.13, \citealp{2003AJ....125.3185P}) , both observed with SDSS.  
The results are shown in Figure \ref{fig:metalratios}.  
Previous studies 
 \citep{2004AJ....128..426W} indicate that subdwarfs are $\sim$ 0.2 mags redder than solar-metallicity stars in $g-r$.
This is most likely due to the multiple hydride bands present in the $g$ 
filter 
\citep{1977ApJ...214..778H,1995bmsb.conf..239D, 2004AJ....128..426W}.
The region between 4000-4500\AA\ in Figure \ref{fig:metalratios}
(left panel) shows that the flux in the subdwarf is depressed by $\sim$ 60\% 
compared to the level present in the composite template.  Strong hydride bands, such 
as MgH near 5000\AA\ and CaH bands near 6800\AA\ are also depressed.  These bands
are labeled in 
Figure \ref{fig:spec}.

In contrast, the metal-rich Hyades star (right panel of Fig. \ref{fig:metalratios})
shows mostly enhanced but variable continuum in the $g$ band, which is
difficult to attribute to any particular feature.  There is enhanced
emission in the core of Na \rm{I} D and Ca II H and K, but these are likely
not strong enough to influence the colors.  Unfortunately we do not
have colors in the SDSS filters, measured with the 2.5m SDSS telescope,
for the Hyades stars, and therefore cannot directly compare the spectral
features with measured color differences between the metal-rich stars
and our templates.


\subsection{Photometric Differences: Colors}\label{sec:photcolors}
Photometry was obtained from the SDSS CAS for each star used in constructing the templates. The average colors for each template are
listed in Table \ref{table:colors} by spectral type.  Previous studies have been inconclusive,
suggesting that active stars are marginally bluer in $U-B$ \citep{1997A&A...319..967A},
redder in $V-I$ \citep{1999AJ....117.1341H} or not statistically different
from inactive populations for SDSS colors \citep{2004AJ....128..426W}.
We computed the color difference for each spectral subtype (active color - inactive color) and
averaged over spectral type.   
The data in Table \ref{table:colors} indicate the following general trends:  active stars
are $\sim 0.09 \pm 0.24$ mags bluer than their non-active counterparts in $u-g$, while they are 
$\sim 0.05 \pm 0.03$ mags redder in $i-z$.   We note that while these trends are suggestive, but are within the
scatter.  No strong trends were present in $g-r$ or $r-i$.

Our goal was to link changes in the spectral features to differences in photometric colors.
Due to the spectral coverage of the SDSS spectra, we are only able to
investigate the
$g-r$ and $r-i$ colors in detail, which did not demonstrate any discernible trends with activity.  The bluer $g-r$ color in the M0 active template
appears to be anomalous, and from the spectral analysis, may be due to low
level flaring - see 
\S \ref{sec:specfeatures}.  This may also be simply 
due to the small number of spectra associated with the active M0 template.  However,
the photometric trend in $u-g$, where active stars were an average of  $\sim$ 0.09 mags bluer than inactive stars, may be reflecting the presence of
similar low level flaring in many of the active templates, as the
enhanced blue continuum during flares will appear even more strongly in the $u$ band
\citep{1976ApJS...31...61M,1991ApJ...378..725H}.  

The small number of M subdwarfs identified in the SDSS database, and
the lack of SDSS photometry for the Hyades M dwarfs prevented us from
investigating color differences due to metallicity.  As noted in the 
previous section, \cite{2004AJ....128..426W} showed that M subdwarfs 
are $\sim$ 0.2 mags redder in $g-r$ than their solar-metallicity 
counterparts.

\section{Conclusions}\label{sec:conclusions}
We used the large SDSS spectral database from DR3 to form active, inactive and
composite template spectra of M dwarfs spanning types M0-L0, 
on a uniform, zero-velocity scale.
Our spectral templates provide suitable radial velocity standards
for analyzing spectra with R$\sim$ 1,800, with
an external accuracy of 
 3.8 km s$^{\rm -1}$, within the quoted error associated with the 
wavelength scale
for SDSS spectroscopy \citep{2000AJ....120.1579Y}.  Internally, the templates 
are consistent to $<$ 1 km s$^{\rm -1}$.

The magnetically active templates, as identified by the presence of 
H$\alpha$ emission in the individual stellar spectra, showed 
little difference in
the measured Balmer decrements with spectral type, indicating that
chromospheric structure and heating are apparently similar
through the M dwarf sequence.  Flares cause much larger changes
in the decrement.  We found some evidence that color changes
(active stars appearing bluer in $u-g$ and in one case in $g-r$)
are due primarily to blue continuum enhancements in the active
stars, which may be due to intermittent low-level flaring. 
In general, chromospheric line emission has a negligible effect
on the colors of active stars.  Molecular bands including 
TiO2, TiO4 and CaOH showed significant changes between the
active and inactive templates.

With regard to metallicity, our findings extend the 
earlier study by \cite{2004AJ....128..426W}, which found
subdwarfs to be $\sim$ 0.2 mags redder in $g-r$.  
Our spectral analysis shows that the flux in the blue is depressed
by as much as 60\%, and that the strong MgH and
CaH bands are significantly deeper in the subdwarfs. 
  The spectral analysis of metal-rich Hyades stars
([Fe/H] = 0.13, \citealp{2003AJ....125.3185P}) showed continuum differences, but these were not
obviously attributed to any particular features.  

The authors would like to thank Andrew Becker and Kelle Cruz for their enlightening conversations.
The authors gratefully acknowledge the support of NSF grant AST02-05875 and NASA ADP grant NAG5-13111.
This research has made use of NASA's Astrophysics Data System Abstract Service,
the SIMBAD database, operated at CDS, Strasbourg, France.  This project made extensive use of SDSS data.
Funding for the SDSS and SDSS-II has been provided by the Alfred P. Sloan Foundation, the Participating Institutions, the National Science Foundation, the U.S. Department of Energy, the National Aeronautics and Space Administration, the Japanese Monbukagakusho, the Max Planck Society, and the Higher Education Funding Council for England. The SDSS Web Site is http://www.sdss.org/.

The SDSS is managed by the Astrophysical Research Consortium for the Participating Institutions. The Participating Institutions are the American Museum of Natural History, Astrophysical Institute Potsdam, University of Basel, Cambridge University, Case Western Reserve University, University of Chicago, Drexel University, Fermilab, the Institute for Advanced Study, the Japan Participation Group, Johns Hopkins University, the Joint Institute for Nuclear Astrophysics, the Kavli Institute for Particle Astrophysics and Cosmology, the Korean Scientist Group, the Chinese Academy of Sciences (LAMOST), Los Alamos National Laboratory, the Max-Planck-Institute for Astronomy (MPIA), the Max-Planck-Institute for Astrophysics (MPA), New Mexico State University, Ohio State University, University of Pittsburgh, University of Portsmouth, Princeton University, the United States Naval Observatory, and the University of Washington. 


\clearpage
\begin{deluxetable}{cll}
 \tablecaption{DAS Query Color Ranges}
 \tablewidth{2.8in}
 \tablehead{
 \colhead{Sp. Type}&
 \colhead{$r-i$}&
 \colhead{$i-z$}
 }
 \startdata
 M0 &  0.50 - 0.85 & 0.30 - 0.50 \\
 M1 &  0.60 - 1.15 & 0.30 - 0.65 \\
 M2 &  0.80 - 1.30 & 0.40 - 0.75 \\
 M3 &  0.90 - 1.50 & 0.40 - 0.90 \\
 M4 &  1.10 - 1.80 & 0.60 - 1.10 \\
 M5 &  1.45 - 2.20 & 0.80 - 1.15 \\
 M6 &  1.65 - 2.25 & 0.90 - 1.20 \\
 M7 &  1.90 - 2.70 & 0.95 - 1.65 \\
 M8 &  2.65 - 2.85 & 1.20 - 1.90 \\
 M9 &  2.85 - 3.05 & 1.25 - 1.80\\
 L0 &   2.30 - 2.70 & 1.70 - 1.90\\
 \enddata
 \label{table:colorcuts}
 
\end{deluxetable}

\begin{deluxetable}{cllllll}

 \tabletypesize{\scriptsize}

 \tablewidth{6in}
 \tablecaption{Active Template Decrements}

\tablehead{
\colhead{Sp. Type}&
\colhead{H$\alpha$}&
\colhead{H$\beta$}&
\colhead{H$\gamma$}&
\colhead{H$\delta$}&
\colhead{H$8$}&
\colhead{Ca {\rm II} K}
}
\startdata

M0 & 2.09 (0.22) & 1.00 (0.15) & \nodata & \nodata & \nodata & \nodata \\
M1 & 2.33 (0.50) & 1.00 (0.28) & \nodata & \nodata & \nodata & 0.19 (0.16) \\
M2 & \nodata & \nodata & \nodata & \nodata & \nodata & \nodata \\
M3 & 3.08 (0.73) & 1.00 (0.28) & \nodata & \nodata & \nodata & \nodata \\
M4 & 3.37 (0.67) & 1.00 (0.26) & 0.51 (0.26) & 0.43 (0.25) & \nodata & \nodata \\
M5 & 4.27 (1.33) & 1.00 (0.34) & 1.12 (0.41) & 0.52 (0.30) & 0.13 (0.36) & 0.86 (0.38) \\
M6 & 3.68 (0.66) & 1.00 (0.24) & 0.76 (0.25) & 0.36 (0.22) & 0.31 (0.33) & 0.67 (0.27) \\
M7 & 4.18 (0.71) & 1.00 (0.23) & 0.72 (0.23) & 0.47 (0.23) & 0.25 (0.32) & 0.80 (0.27) \\
M8 & 5.90 (1.11) & 1.00 (0.25) & 0.90 (0.28) & 0.64 (0.28) & 0.32 (0.39) & \nodata \\
M9 & \nodata & \nodata & \nodata & \nodata & \nodata & \nodata \\
L0 & \nodata & \nodata & \nodata & \nodata & \nodata & \nodata \\

\hline
Average\tablenotemark{a} & 3.61 (1.21) & 1.00 (0.00) & 0.80 (0.23) & 0.48 (0.11) & 0.25 (0.09) & 0.78 (0.10) \\
AD Leo \citep{1991ApJ...378..725H} & \nodata &  1.00 &   0.81 &     0.69 &     0.50  & 0.18\\ 
Quiet Model \citep{2006ApJ...644..484A} &   2.20 &      1.00 &     0.58 &     0.21 &     0.34 & 3.65\\
Flare Model \citep{2006ApJ...644..484A} &  0.57 &      1.00 &     0.84 &     0.75 &     0.54 & 0.10\\
\enddata
\label{table:decrements}
\tablecomments{Decrement measurements are reported with measurement errors in parentheses.}
\tablenotetext{a}{Errors reported on means are 1 $\sigma$ of individual decrement measurements.}
\end{deluxetable}

\begin{deluxetable}{crrrrrrr}

 \tabletypesize{\scriptsize}

 \tablecaption{Active Template Equivalent Widths and $L_{\rm H\alpha}/L_{\rm bol, {\it i-z}}$}

\tablehead{
\colhead{Sp. Type}&
\colhead{H$\alpha$ EW}&
\colhead{H$\beta$ EW}&
\colhead{H$\gamma$ EW}&
\colhead{H$\delta$ EW}&
\colhead{H$8$ EW}&
\colhead{Ca {\rm II} K EW}&
\colhead{$L_{\rm H\alpha}/L_{\rm bol, {\it i-z}}$}
}
\startdata
M0 & 1.39 (0.04)  & 1.09 (0.11)  & \nodata  & \nodata  & \nodata & \nodata &  2.15E-04 (5.48E-05) \\
M1 & 1.33 (0.10)  & 1.54 (0.31)  & \nodata  & \nodata  & \nodata & 1.03 (0.84) & 1.28E-04 (5.83E-05) \\ 
M2 & 3.57 (0.10)  & \nodata  & \nodata  & \nodata  & \nodata & \nodata &  3.24E-04 (2.87E-05) \\
M3 & 2.45 (0.31)  & 2.40 (0.48)  & \nodata  & \nodata & \nodata & \nodata & 1.42E-04 (7.00E-05)  \\
M4 & 4.12 (0.35)  & 5.10 (0.93)  & 6.33 (3.09)  & 7.17 (4.09)  & \nodata & \nodata & 1.75E-04 (6.08E-05)  \\
M5 & 5.85 (1.16)  & 5.56 (1.34)  & 16.15 (4.56)  & 8.28 (4.41)  & 3.78 (10.26) & 25.09 (12.75) & 1.82E-04 (7.14E-05)  \\
M6 & 6.06 (0.39)  & 7.94 (1.34)  & 18.88 (5.55)  & 9.36 (5.75)  & 9.26 (10.20) & 19.10 (9.16) & 1.35E-04 (2.43E-05)  \\
M7 & 8.15 (0.50)  & 10.47 (1.70)  & 25.85 (7.89)  & 15.73 (7.64)  & 10.29 (13.44) & 21.30 (8.84) & 1.01E-04 (3.45E-05) \\
M8 & 10.99 (0.64)  & 12.23 (2.24)  & 113.05 (60.46)  & 38.65 (19.74)  & 20.34 (27.37) & \nodata & 4.41E-05 (1.14E-05)  \\
M9 & 6.10 (0.18)  & \nodata  & \nodata  &  \nodata  & 10.94 (14.52) &  \nodata & 2.52E-05 (6.41E-06)  \\
L0  & 6.86 (0.41)  & \nodata & \nodata & \nodata  & \nodata &  \nodata &  1.97E-05 (2.50E-06) \\

\enddata

\label{table:ew}
\tablecomments{Equivalent Widths are reported in \AA with measurement errors in parentheses.\\
$L_{\rm H\alpha}/L_{\rm bol, {\it i-z}}$ measurement errors are also in parentheses.}
\end{deluxetable}

\clearpage
 
 \begin{deluxetable}{c|rrr|rrr|rrr}

\tablewidth{7in}
\tabletypesize{\scriptsize}

 \tablecaption{Template Bandheads}

 \tablehead{
\colhead{Sp. Type}&
\multicolumn{3}{c}{CaH1}&
\multicolumn{3}{c}{CaH2}&
\multicolumn{3}{c}{CaH3}
}

\startdata
	& 	Active	& 	Inactive 	& 	All &	Active	& 	Inactive 	& 	All &	Active	& 	Inactive 	& 	All \\

M0 & 0.94 (0.00) & 0.91 (0.01) & 0.91 (0.01) & 0.79 (0.00) & 0.79 (0.01) & 0.79 (0.01) & 0.87 (0.00) & 0.89 (0.01) & 0.89 (0.01)\\
M1 & 0.86 (0.01) & 0.85 (0.02) & 0.85 (0.02) & 0.68 (0.01) & 0.67 (0.01) & 0.67 (0.01) & 0.82 (0.01) & 0.83 (0.02) & 0.83 (0.02)\\
M2 & 0.81 (0.01) & 0.80 (0.03) & 0.80 (0.03) & 0.49 (0.01) & 0.56 (0.03) & 0.56 (0.03) & 0.68 (0.01) & 0.76 (0.04) & 0.76 (0.04)\\
M3 & 0.74 (0.04) & 0.78 (0.07) & 0.78 (0.07) & 0.48 (0.03) & 0.48 (0.06) & 0.48 (0.06) & 0.71 (0.04) & 0.72 (0.09) & 0.72 (0.09)\\
M4 & 0.76 (0.03) & 0.76 (0.04) & 0.76 (0.04) & 0.39 (0.02) & 0.43 (0.02) & 0.42 (0.02) & 0.65 (0.03) & 0.69 (0.04) & 0.68 (0.04)\\
M5 & 0.75 (0.04) & 0.80 (0.03) & 0.78 (0.04) & 0.37 (0.03) & 0.38 (0.02) & 0.38 (0.02) & 0.64 (0.04) & 0.67 (0.03) & 0.66 (0.04)\\
M6 & 0.76 (0.03) & 0.79 (0.07) & 0.77 (0.05) & 0.33 (0.01) & 0.33 (0.03) & 0.33 (0.02) & 0.60 (0.02) & 0.64 (0.06) & 0.62 (0.04)\\
M7 & 0.78 (0.04) & 0.77 (0.04) & 0.78 (0.04) & 0.29 (0.01) & 0.28 (0.01) & 0.28 (0.01) & 0.58 (0.02) & 0.59 (0.02) & 0.58 (0.02)\\
M8 & 0.84 (0.04) & \nodata & 0.85 (0.04) & 0.28 (0.01) & \nodata  & 0.28 (0.01) & 0.57 (0.01) & \nodata & 0.57 (0.01) \\
M9 & 0.90 (0.03) & \nodata & 0.91 (0.03) & 0.30 (0.01) & \nodata  & 0.30 (0.01) & 0.63 (0.01) & \nodata & 0.63 (0.01) \\
L0 & 0.97 (0.04) & \nodata & 0.96 (0.04) & 0.50 (0.01) & \nodata  & 0.50 (0.01) & 0.71 (0.01) & \nodata & 0.71 (0.01) \\

\enddata
\label{table:cah}
\tablecomments{Measurement errors are given in parentheses.}
\end{deluxetable}

\begin{turnpage}
 \begin{deluxetable}{c|rrr|rrr|rrr|rrr}

\tablewidth{9.5in}
\tabletypesize{\scriptsize}

 \tablecaption{Template Bandheads}

 \tablehead{
\colhead{Sp. Type}&
\multicolumn{3}{c}{TiO2}&
\multicolumn{3}{c}{TiO4}&
\multicolumn{3}{c}{TiO5}&
\multicolumn{3}{c}{TiO8}
}
\startdata
	&  Active	& 	Inactive 	& 	All &	Active	& 	Inactive 	& 	All &	Active	& 	Inactive 	& 	All &	Active	& 	Inactive 	& 	All \\
M0 & 0.86 (0.01) & 0.91 (0.02) & 0.91 (0.02) & 0.89 (0.01) & 0.91 (0.02) & 0.91 (0.02) & 0.78 (0.01) & 0.82 (0.01) & 0.82 (0.01)& 0.97 (0.00) & 0.99 (0.01) & 0.99 (0.01) \\
M1 & 0.79 (0.03) & 0.85 (0.03) & 0.84 (0.03) & 0.83 (0.02) & 0.85 (0.02) & 0.85 (0.02) & 0.71 (0.01) & 0.72 (0.02) & 0.72 (0.02)& 0.98 (0.01) & 0.98 (0.01) & 0.98 (0.01) \\
M2 & 0.72 (0.02) & 0.78 (0.05) & 0.77 (0.05) & 0.75 (0.02) & 0.79 (0.05) & 0.79 (0.05) & 0.52 (0.01) & 0.61 (0.03) & 0.60 (0.03)& 0.97 (0.01) & 0.97 (0.03) & 0.97 (0.03) \\
M3 & 0.65 (0.07) & 0.69 (0.13) & 0.69 (0.13) & 0.69 (0.06) & 0.70 (0.10) & 0.70 (0.10) & 0.49 (0.03) & 0.49 (0.07) & 0.49 (0.07)& 0.99 (0.03) & 0.92 (0.06) & 0.93 (0.06) \\
M4 & 0.56 (0.04) & 0.62 (0.05) & 0.61 (0.05) & 0.64 (0.04) & 0.64 (0.04) & 0.64 (0.04) & 0.39 (0.02) & 0.41 (0.03) & 0.41 (0.02)& 0.91 (0.02) & 0.90 (0.03) & 0.90 (0.02) \\
M5 & 0.51 (0.05) & 0.54 (0.04) & 0.53 (0.05) & 0.60 (0.04) & 0.60 (0.04) & 0.60 (0.04) & 0.34 (0.03) & 0.34 (0.02) & 0.34 (0.02)& 0.84 (0.03) & 0.84 (0.02) & 0.84 (0.03) \\
M6 & 0.43 (0.02) & 0.46 (0.07) & 0.44 (0.05) & 0.56 (0.03) & 0.54 (0.07) & 0.55 (0.05) & 0.28 (0.01) & 0.27 (0.03) & 0.27 (0.02)& 0.77 (0.01) & 0.77 (0.04) & 0.77 (0.03) \\
M7 & 0.35 (0.02) & 0.37 (0.02) & 0.35 (0.02) & 0.53 (0.03) & 0.47 (0.03) & 0.51 (0.03) & 0.22 (0.01) & 0.20 (0.01) & 0.22 (0.01)& 0.68 (0.01) & 0.68 (0.01) & 0.68 (0.01) \\
M8 & 0.32 (0.02) & \nodata & 0.32 (0.02) & 0.65 (0.03) & \nodata & 0.65 (0.03) & 0.25 (0.01) & \nodata & 0.25 (0.01) & 0.55 (0.01) & \nodata & 0.55 (0.01) \\
M9 & 0.32 (0.01) & \nodata & 0.32 (0.01) & 0.62 (0.02) & \nodata & 0.61 (0.02) & 0.26 (0.01) & \nodata & 0.26 (0.01) & 0.54 (0.00) & \nodata & 0.54 (0.00) \\
L0 & 0.65 (0.04) & \nodata & 0.64 (0.04) & 0.87 (0.05) & \nodata & 0.86 (0.05) & 0.64 (0.02) & \nodata & 0.64 (0.02) & 0.62 (0.01) & \nodata & 0.62 (0.01) \\

\enddata
\label{table:tio}
\tablecomments{Measurement errors are given in parentheses.}
\end{deluxetable}

\clearpage{\thispagestyle{empty}}

 \begin{deluxetable}{c|rrr|rrr|rrr|rrr|rrr}

\tablewidth{9.5in}
\tabletypesize{\scriptsize}
\setlength{\tabcolsep}{0.04in}
\tablecaption{Template Colors}

 \tablehead{
\colhead{Sp. Type}&
\multicolumn{3}{c}{Num. Stars \tablenotemark{a}}&
\multicolumn{3}{c}{$u-g$}&
\multicolumn{3}{c}{$g-r$}&
\multicolumn{3}{c}{$r-i$}&
\multicolumn{3}{c}{$i-z$}
}
\startdata
	&  &	 & &	Active	& 	Inactive	&	All	&	Active	&	Inactive	&	All	&	Active	& 	Inactive 	& 	All &	Active	& 	Inactive 	& 	All \\
	
M0 & 7 & 570 & 577 & 2.21 (0.40) & 2.56 (0.66) & 2.56 (0.66) & 1.24 (0.38) & 1.40 (0.55) & 1.40 (0.54) & 0.68 (0.13) & 0.65 (0.12) & 0.65 (0.12) & 0.41 (0.09) & 0.39 (0.08) & 0.39 (0.08) \\
M1 & 4 & 275 & 279 & 2.17 (0.51) & 2.54 (0.64) & 2.53 (0.64) & 1.43 (0.19) & 1.47 (0.44) & 1.47 (0.44) & 0.77 (0.16) & 0.80 (0.11) & 0.80 (0.11) & 0.58 (0.16) & 0.47 (0.07) & 0.47 (0.07) \\
M2 & 1 & 66 & 67 & 2.69 (0.03) & 2.29 (0.80) & 2.30 (0.80) & 1.90 (0.03) & 1.60 (0.36) & 1.60 (0.35) & 1.07 (0.03) & 1.04 (0.18) & 1.04 (0.18) & 0.60 (0.03) & 0.59 (0.11) & 0.59 (0.11) \\
M3 & 6 & 186 & 192 & 2.00 (0.35) & 2.18 (0.85) & 2.18 (0.84) & 1.69 (0.18) & 1.59 (0.23) & 1.60 (0.22) & 1.30 (0.15) & 1.28 (0.19) & 1.28 (0.19) & 0.76 (0.17) & 0.70 (0.12) & 0.70 (0.12) \\
M4 & 25 & 137 & 162 & 2.28 (1.09) & 2.28 (0.77) & 2.28 (0.82) & 1.60 (0.17) & 1.55 (0.21) & 1.56 (0.20) & 1.49 (0.28) & 1.42(0.16) & 1.43 (0.19) & 0.87 (0.12) & 0.79 (0.10) & 0.81 (0.10) \\
M5 & 171 & 235 & 406 & 2.13 (0.95) & 2.18 (0.89) & 2.16 (0.92) & 1.52 (0.30) & 1.57 (0.24) & 1.55 (0.27) & 1.74 (0.21) & 1.72 (0.21) & 1.73 (0.21) & 0.98 (0.12) & 0.95 (0.11) & 0.96 (0.12) \\
M6 & 1132 & 899 & 2031 & 2.12 (0.90) & 2.17 (0.93) & 2.15 (0.91) & 1.59 (0.13) & 1.55 (0.13) & 1.57 (0.13) & 1.98 (0.11) & 1.99 (0.12) & 1.98 (0.12) & 1.10 (0.06) & 1.09 (0.06) & 1.09 (0.06) \\
M7 & 400 & 150 & 550 & 1.89 (0.91) & 1.98 (0.93) & 1.92 (0.92) & 1.63 (0.17) & 1.55 (0.15) & 1.61 (0.17) & 2.33 (0.19) & 2.35 (0.18) & 2.34 (0.19) & 1.31 (0.12) & 1.26 (0.09) & 1.29 (0.11) \\
M8 & 16 & \nodata & 16 & 1.65 (0.93) & \nodata & 1.65 (0.93) & 1.80 (0.16) & \nodata & 1.80 (0.16) & 2.76 (0.12) & \nodata &2.76 (0.12) & 1.71 (0.09) & \nodata & 1.71 (0.09) \\
M9 & 5 & \nodata & 5 & 1.79 (0.79) & \nodata & 1.79 (0.79) & 1.74 (0.14) & \nodata & 1.74 (0.14) & 2.83 (0.07) & \nodata & 2.83 (0.07) & 1.70 (0.09) & \nodata & 1.70 (0.09) \\
L0 & 4 & \nodata & 4 & 1.35 (1.35) & \nodata & 1.35 (1.35) & 2.50 (0.39) & \nodata & 2.50 (0.39) & 2.54 (0.08) & \nodata & 2.54 (0.08) & 1.83 (0.04) & \nodata & 1.83 (0.04) \\
	\enddata
\label{table:colors}
\tablenotetext{a}{Number of stars composing each template spectrum.}
\tablecomments{Mean SDSS colors are reported in magnitudes, with the one $\sigma$ spread at each 
spectral type reported in parentheses.}
\end{deluxetable}

\end{turnpage}

\end{document}